\documentclass[intlimits,twoside,a4paper]{article}

\usepackage{amsmath,amssymb}
\usepackage{graphicx}
\usepackage[T2A]{fontenc}
\usepackage[cp1251]{inputenc}

\usepackage[eqsecnum]{cmpj2}

\usepackage{color}

\issue{2016}{19}{3}{33601}
\doinumber{10.5488/CMP.19.33601}

\title[Cylindrically confined assembly of diblock copolymer under oscillatory shear flow]
{Cylindrically confined assembly of diblock copolymer under oscillatory shear flow}

\author[Y.-Q.~Guo \textsl{et al.}]
{Y.-Q.~Guo, J.-X.~Pan, J.-J.~Zhang, M.-N.~Sun, B.-F.~Wang, H.-Sh.~Wu}

\address{School of Chemistry and Materials Science, Shanxi Normal University, Linfen, 041004, China}

\authorcopyright{Y.-Q.~Guo, J.-X.~Pan, J.-J.~Zhang, M.-N.~Sun, B.-F.~Wang, H.-Sh.~Wu, 2016}
\date{Received November 30, 2015, in final form January 18, 2016}

\begin{document}

\maketitle

\begin{abstract}

Manipulating the self-assembly nanostructures with combined
different control measures is emerging as a promising route for
numerous applications to generate templates and scaffolds for
nanostructured materials. Here, the two different control measures
are a cylindrical confinement and an oscillatory shear flow. We
study the phase behavior of diblock copolymer confined in nanopore
under oscillatory shear by considering different $D/L_0$ ($D$ is
the diameter of the cylindrical nanopore, $L_0$ is the domain
spacing) and different shears via Cell Dynamics Simulation. Under
different $D/L_0\,$, in the system occurs different morphology
evolution and phase transition with the changing of amplitude and
frequency. Meanwhile, it forms a series of novel morphologies. For
each $D/L_0\,$, we construct a phase diagram of different forms and
analyze the reason why the phase transition occurs. We find that
although the morphologies are different under different $D/L_0\,$,
the reason of the phase transition with the changing of amplitude
and frequency is roughly the same, all caused by the interplay of
the field effect and confinement effect. These results can guide
an experimentalist to an easy method of creating the ordered,
defect-free nanostructured materials using a combination of the
cylindrical confinement and oscillatory shear flow.

\pacs 64.75.Yz, 64.75.Va, 83.80.Uv, 83.10.Tv
\end{abstract}

\section{Introduction}

Block copolymer is a class of soft material capable of
self-assembling to form ordered structures at nanometer scales.
These structures possess a potential for various applications to
nanotechnologies such as lithographic templates for nanowires,
photonic crystals, high-density magnetic storage media, drug
delivery and biomineralization. So, the self-assembly of block
copolymer has attracted much interest as an efficient and
effective means to create structures at nanometer scales.
Obviously, how to manipulate these self-assembled nanostructures
have become the key of the research, so understanding the
controlling factors is an intriguing field of research that holds
promise for further expanding the nanofabrication. In general, the
control measures mainly include a confined environment \cite{a,b},
imposed external field \cite{c,d}, substrate induction \cite{e,f},
doping \cite{g,h}, and branched molecular architectures \cite{h1}
etc. Thereinto, confinement effect makes the polymer system form a
series of novel morphologies that are not accessible in the bulk,
and the external effect makes the structures more ordered. They
both provide an effective route to manipulate the self-assembled
nanostructures.

During the past years, the effects of a confined environment have
been extensively explored  based on a large number of experimental
and theoretical studies. The internal branching is one of the
confinements, i.e., one imposes some restrictions on the
conformational freedom. For example, Ilnytskyi et~al. analyzed the
peculiarities of the equilibrium morphologies observed for the
star and comb diblock polymers with equal molecular mass but with
the differences in both intramolecular architecture and
composition fraction \cite{h1}. In addition, the geometrical
confinement can be classified as one-, two-, and
three-dimensionally confined systems. The one-dimensionally
confined system \cite{i,j} is the simplest case, and its research
is very mature. For a more complex three-dimensio\-nal\-ly
confined system, the studies are relatively few so far \cite{k,l}.
The research area of two-dimensional confinement is developing
rapidly. When placing the block copolymer in a cylindrical pore, a
series of novel morphologies that were not accessible in the bulk
or in the one-dimensionally confined systems were observed at
different degrees of confinement $D/L_0\,$. Russell and co-workers
discovered that the lamellae-, cylinder-, and sphere-forming block
copolymer of polystyrene-$b$-polybutadiene (PS-$b$-PBD) were
confined in the pores of anodized aluminumoxide (AAO) membranes
experimentally \cite{m,n,o,p,q}. They find that under cylindrical
confinement, block copolymer formed various kinds of novel
structures that were not accessible in the bulk, such as stacked
disks, concentric cylinders, torus-like structure, core-shell
structure, single row and zigzag arrangement of spherical
microdomains, single-, double-, triple-helical structures etc. Sun
et al. also observed the diameter-dependence of morphology
confined within the ordered porous alumina templates, but the
block copolymer that they used was polystyrene-$b$-poly (methyl
methacrylate) (PS-$b$-PMMA) \cite{r} and
polyethylene-co-butylene-$b$-polyethylene oxide (PHB-$b$-PEO)
experimentally \cite{s}. For symmetric lamellae-forming diblock
copolymer in cylindrical pores, many novel morphologies were
predicted based on dynamic density functional theory
(DDFT) \cite{t,u,v}, Monte Carlo (MC) \cite{w,x}, self-consistent
field theory (SCFT) \cite{y}, and simulated annealing technique
(SAT) \cite{z}. For the asymmetric cylinder-forming diblock
copolymer, the structures in the cylindrical pores severely
deviated from the bulk. The novel structures of spontaneous
formation, such as stacked toroids, single helix, double helix and
perforated tubes, were observed by means of a SAT \cite{aa,bb} and
SCFT \cite{cc,dd,ee}. For the asymmetric sphere-forming diblock
copolymer, the study of it was rare. Pinna et al. predicted
tremendous rich morphologies of sphere-forming diblock copolymer
in cylindrical nanopores by using the cell dynamics simulation
(CDS) \cite{ff}. Recently, the phase behavior of sphere-forming
triblock copolymer confined in nanopores was investigated by Hao
et al. by means of a DDFT~\cite{gg}. They observed that typical
structures were different from the bulk morphologies, which were
consistent with the available experiments.

As concerns the imposition of an external field, this is a
versatile control measure to obtain the long-range order and then
to create microstructures with potential applications in
biomaterials, optics, and microelectronics. Earlier, during the
study of block copolymer there were observed various alignments in
a lamellar block copolymer by using TEM and SAXS. Thereinto,
polystyrene-polyisoprene is the most representative. The parallel
and perpendicular orientations were observed by K.I.~Winey and S.S.~Patel et~al. in 1993~\cite{hh} and 1995~\cite{ii},
respectively; the transverse orientation was found by V.K.~Gupta
et~al.~\cite{jj} and Y.M.~Zhang~\cite{kk}. Then, B.S.~Pinheiro
and K.I.~Winey observed the mixed parallel-perpendicular
morphologies in diblock copolymer systems at intermediate
temperatures~\cite{ll}. Afterwards, researchers theoretically
predicted the phase transition of the lamellar~\cite{mm,nn,oo,pp},
ring~\cite{qq}, hexagonal cylinders~\cite{rr} diblock copolymer
subjected to shear flow further. These predictions were proved by
researchers with various numerical simulation methods, such as
CDS~\cite{ss,tt,uu,vu}, nonequilibrium molecular dynamics
simulation (NEMD)~\cite{vv,ww}, dissipative particle dynamics
method (DPD)~\cite{xx,yy}, SCFT and lattice Boltzmann (LB)
method~\cite{zz}, Brownian dynamics (BD)~\cite{ab}, mean-field
approach (MFA)~\cite{cd}, density functional theory
(DFT)~\cite{ef,gh}, MC~\cite{ij} etc. For all these cases, shear
flow plays an important role as a means for aligning the
microscopic domains.

As we all know, the polymer system can form all kinds of novel
structures that are different from the bulk morphologies under
cylindrical confinement and form ordered structures under shear
flow. So, we wonder what the phase behavior of polymer system is
under both these two control measures, and it should be
interesting when these two control measures affect the system.
Previously, Pinna et~al.~\cite{kl}  combined the two control
measures of the one-dimension confinement and steady shear flow.
They demonstrated the shear alignment and the shear-induced
transitions in sphere-forming diblock copolymer single layer and
bilayer films by cell dynamics simulation that was observed
experimentally by Hong et~al.~\cite{mn}, and for the first time
presented a nontrivial alignment mechanism of a single layer of
spherical domains in shear. On this basis, we dedicate ourselves
to investigate the phase behavior of diblock copolymer within
cylindrical pores under oscillatory shear flow by means of the CDS
of time dependent Ginzbrug-Landau (TDGL) theory proposed by Oono
and co-workers~\cite{op,qr,st,uv,wx}. It will further provide some
guiding function for experimentalists. The other parts of this
paper are organized as follows: section \ref{Model} is devoted to the
description of the model and simulation method; section \ref{Result} is the
numerical results and discussions; and finally, section \ref{Conc} gives a
brief conclusion in this work.

\newpage
\section{Models and simulation methods}\label{Model}

Our simulations are performed in a neutral cylindrical nanopore
with a diameter $D$ and a length $L_z\,$. We employ a TDGL approach
in the form of the CDS, which is a very fast computational
technique. For AB diblock copolymer, the structure can be
described by an order parameter $\phi(\textbf{r},t)=\phi_{\text{A}}-\phi_{\text{B}}+(1-2f)$, where $\phi_{\text{A}}$ and $\phi_{\text{B}}$ are the
local volume fractions of A and B monomers, respectively. The
fraction of A monomers in a diblock copolymer chain is denoted by
$f$, $f=N_{\text{A}}/(N_{\text{A}}+N_{\text{B}})$. The kinetics and morphological evolution
are described, in the spirit of linear irreversible
thermodynamics, by the TDGL equation for a diffusive field coupled
with an external velocity field, which can be written as~\cite{yz}
\begin{equation}
\label{1}
\frac{\partial\phi(\textbf{r},t)}{\partial t}+ \textbf{v} \cdot \pmb\nabla\phi(\textbf{r},t)=M\nabla^2\left\{\frac{\delta
F[\phi(\textbf{r},t)]}{\delta\phi(\textbf{r},t)}\right\},
\end{equation}
where $M$ is a phenomenological mobility constant and is set to 1.
$\textbf{r}$ is an external velocity field.

For simplicity, we set the shear rate as
\begin{equation}
\nu(\textbf{r},t)=\big(0,\, 0,\, \gamma \omega y \cos(\omega t)\big),
\end{equation}
where $\gamma$ is amplitude, and $\omega$ is frequency.
We set $z$-axis (the direction of the arrows) is the shear direction.

Here, we use a two-order-parameter model in  \cite{a1,a2}. The
long-range part $F_{\text{L}}$ and the short-range part $F_{\text{S}}$ are given by
\begin{equation}
F_{\text{L}}=\frac{\alpha}{2}\iint {\rm d} \textbf{r}{\rm d}\textbf{r}' G(\textbf{r},\textbf{r}')[\phi(\textbf{r})-\phi_0][\phi(\textbf{r})-\phi_0],
\end{equation}
and
\begin{equation}
F_{S}=\iint{\rm d}\textbf{r}\left\{f(\phi)+\frac{D}{2}\left[\pmb\nabla\phi(\textbf{r})\right]^2\right\},
\end{equation}
respectively. The long-range part is relatively simple, in which $G(\textbf{r},\textbf{r}')$
is the Green's function defined by the equation
$-\nabla^{2}G(\textbf{r}, \textbf{r}')=\delta(\textbf{r}-\textbf{r}')$,
while $\alpha$ is a parameter that introduces a chain-length dependence to the free-energy, $\phi_0$
is the spatial averages of $\phi$, we should set $\phi_0=0$ in the case of symmetric copolymers.
As for the short-range part, $D$ is a positive constant that plays the role of a diffusion coefficient,
and $f(\phi)=-A \tanh(\phi)+\phi$.
We carry out computer simulations of the model system by using the CDS proposed by Oono and
Puri~\cite{op,qr,st,uv,wx}. In three-dimensional CDS, the system is discretized on a $L_x \times L_y \times L_z$
cubic lattice, and the order parameter for each cell is defined as $\phi(\textbf{n},t)$,
where $\textbf{n}=(n_x\,,n_y\,,n_z)$ is the lattice position and $n_x\,$, $n_y\,$, and $n_z$ are integers
between 1 and $L$. The Laplacian in CDS is approximated by
\begin{equation}
\nabla^2\phi(\textbf{n})=\langle\langle\phi(
\textbf{n})\rangle\rangle-\phi(\textbf{n}),
\end{equation}
where $\langle\langle\phi(\textbf{n})\rangle\rangle$ represents the following summation of $\phi(\textbf{n})$
for the nearest neighbors (n.), the next-nearest neighbors (n.n.), and the next-next-nearest neighbors (n.n.n.)~\cite{wx}
\begin{equation}
\langle\langle\phi(\textbf{n})\rangle\rangle=\frac{6}{80}\sum_{\textbf{n}=\text{n.}}\phi(\textbf{r})+\frac{3}{80}\sum_{\textbf{n}=\text{n.n.}}\phi(\textbf{r})
+\frac{1}{80}\sum_{\textbf{n}=\text{n.n.n.}}
\phi(\textbf{r}).
\end{equation}
In our simulations, we set $\Delta x,\Delta y,\Delta z$ are 1, and
$\Delta t=1.0$. Then, equation~(\ref{1}) can be transformed into the following
difference equation:
\begin{eqnarray}
\phi(\textbf{r},t+\Delta{t})=\phi(\textbf{r},t)-\frac{1}{2}\gamma \sin(\omega t)y t\left[\phi(x,y,z+1,t) -\phi(x,y,z-1,t)\right]+M\left(\langle\langle I_\phi \rangle\rangle-I_\phi\right)-\alpha\phi(\textbf{r},t)
\end{eqnarray}
with
\begin{equation}
    I_\phi=-D\left(\langle\langle\phi\rangle\rangle-\phi\right)-A \tanh({\phi})+\phi.
\end{equation}
We have chosen $z$-axis as the flow direction, $y$-axis as the velocity gradient direction and $x$-axis as the vorticity axis. In addition to Dirichlet boundary condition~\cite{ff}, a shear periodic boundary condition proposed by Ohta et~al.~\cite{a3,a4,a5} has been applied to $z$ direction along the pore. With the shear strain $\Upsilon$, this boundary condition is written as
\begin{equation}
    \phi(n_x\,,n_y\,,n_z\,,t)=\phi\big(n_x+N_xL,\, n_y+N_yL,\, n_z+N_yL+\Upsilon(t)N_yL\big),
\end{equation}
where $N_x\,$, $N_y\,$, and $N_z$ are arbitrary integers. All parameters in this paper are scaled, and all of them are dimensionless~\cite{op}.

\section{Numerical results and discussion}\label{Result}

In order to simulate the confined self-assembly of AB diblock
copolymer under oscillatory shear flow, we construct the
cylindrical nanopore of diameter $D$ in cubic lattice. The volume
is $V=L_x \times L_y \times L_z$ with $L_x=L_y=D+m$ and
$L_{\mathrm{pore}}=L_z\,$, where $m=2$ when $D$ is odd and $m=3$
when $D$ is even. Polymers cannot occupy the wall sites which are
the lattice sites outside the cylinder with diameter $D$ and the
wall is impenetrable. The extra $m$ in $L_x$ and $L_y$ ensures
that each site inside the cylindrical pore can find its nearest
neighbor, the next-nearest neighbor, and the next-next-nearest
neighbor cells. These sites are either inside the pore or in the
wall. Thus, the polymers are confined in the cylindrical nanopore
of diameter $D$. In this paper, we investigate the phase behavior
under oscillatory shear of diblock copolymer in the cylindrical
pore with different $D/L_0\,$, which $L_0$ is the domain spacing,
refer to \cite{ee}.
\begin{figure}[!b]
\begin{center}
\includegraphics[width=0.65\textwidth]{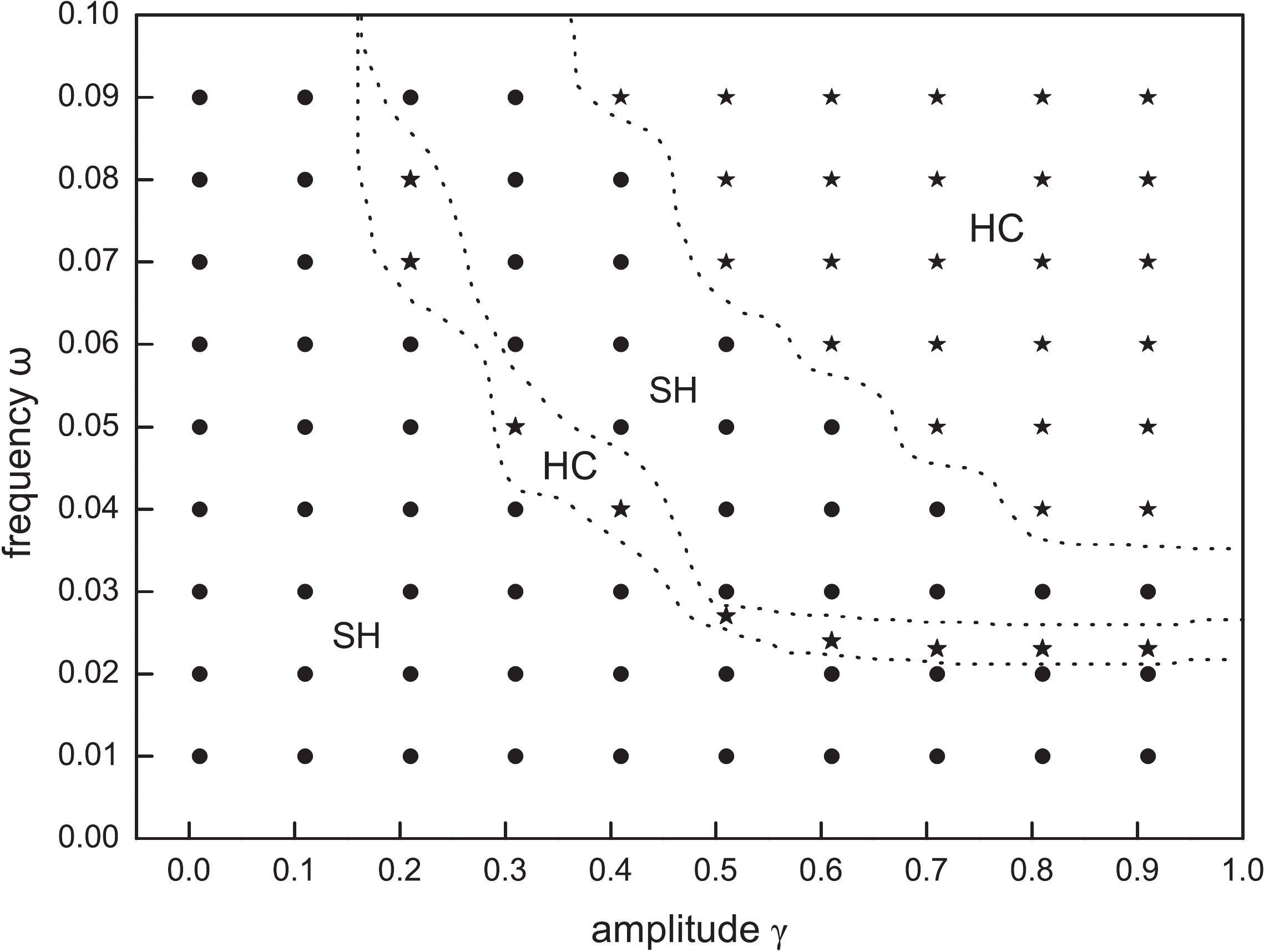}
\end{center}
\caption{Phase diagram of the diblock copolymer in nanopore $D/L_0=1.01$ under oscillatory shear at different amplitudes $\gamma$ and frequencies $\omega$. $\bullet$ represents single-helical structure which is abbreviated as SH; $\star$ represents half-cylinder structure which is abbreviated as HC.} \label{Fig1}
\end{figure}

Firstly, we investigate the effect of oscillatory shear flow in
cylindrical nanopore of $D/L_0=1.01$, $L_{\mathrm{pore}}/L_0=8.1$.
It is noted that the oscillatory shear is imposed on the
disordered structures rather than the final self-assembled
structures. With the simulation on the continuous variation of
shear amplitude and shear frequency, we roughly construct a
structure diagram in figure~\ref{Fig1}. In figure~\ref{Fig1}, symbol $\bullet$ represents
single-helical structure that is abbreviated as SH, $\star$
represents half-cylinder structure that is abbreviated as HC. The
half-cylinder refers to A phase occupying half of the cylinder
along the pipe axis.

\begin{figure}[!t]
\begin{center}
\includegraphics[width=0.9\textwidth]{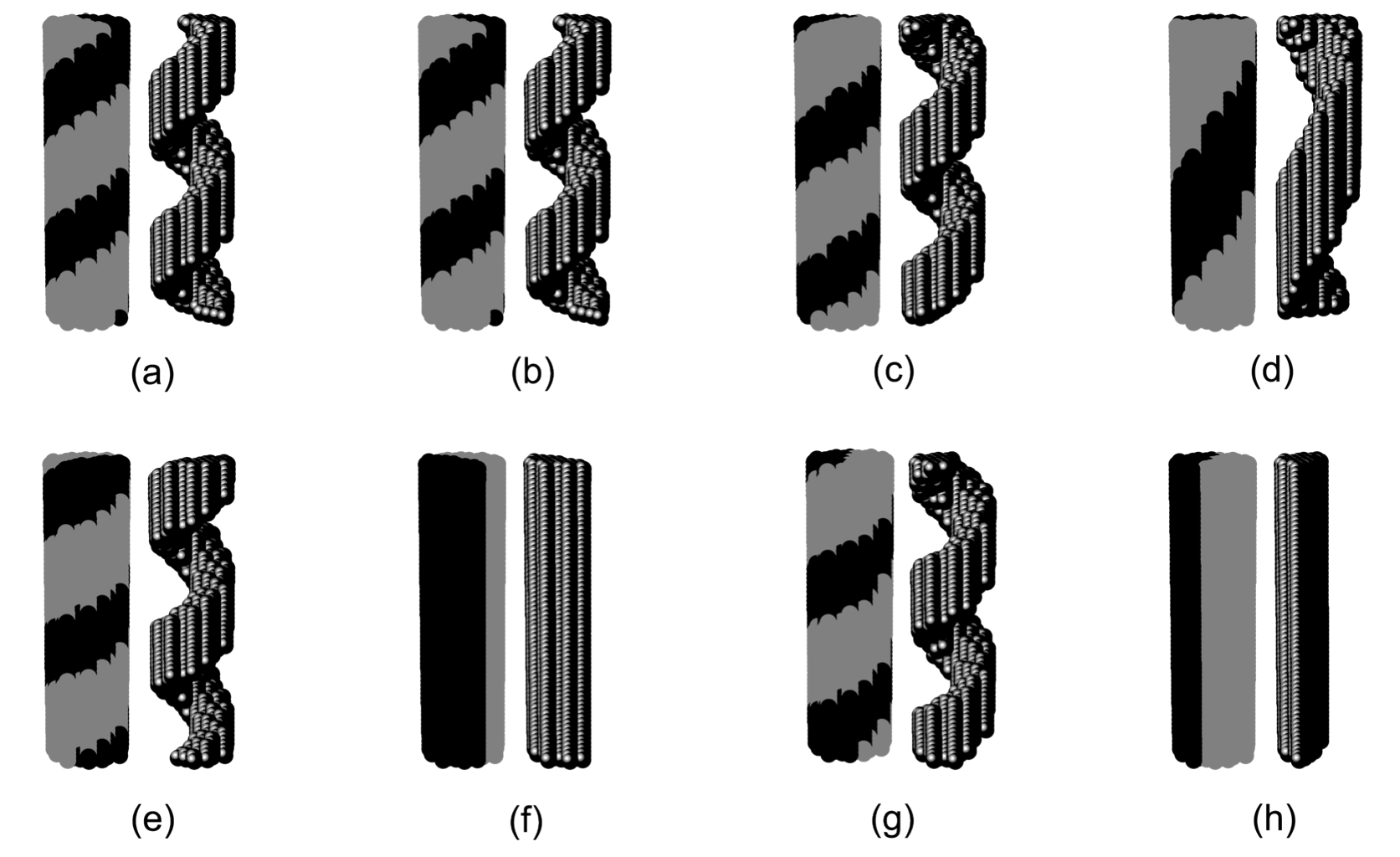}
\end{center}
\vspace{-3mm}
\caption{Pattern evolution of diblock copolymer in nanopore of $D/L_0=1.01$ under different oscillatory shears:  (a) $\gamma=0.01,\,\omega=0.00001$, (b) $\gamma=0.01,\,\omega=0.001$, (c) $\gamma=0.01,\,\omega=0.1$, (d) $\gamma=0.01,\,\omega=1.0$; (e) $\gamma=0.41,\,\omega=0.03$, (f) $\gamma=0.41,\,\omega=0.04$, (g) $\gamma=0.41,\,\omega=0.05$, (h) $\gamma=0.41,\,\omega=0.09$. Phase A is represented by the black regions, phase B by the gray regions.} \label{Fig2}
\end{figure}

In figure~\ref{Fig1}, the amplitude, $\gamma$, varies from 0 to 1.0, and the
frequency, $\omega$, varies from 0 to 0.1. We can clearly see that
in small amplitudes, such as $\gamma\leqslant0.15$, the field effect is
too weak to have a sufficient capability to affect the morphology,
therefore, no phase transition is observed at the given frequency,
and the single-helical structure is always preserved. Typical
snapshots of this case are shown in figures~\ref{Fig2}~(a)--(d). Moreover,
in the small frequencies $\omega\leqslant0.02$, the structure is
single-helical at all amplitudes. At amplitudes $0.2<\gamma<0.35$,
with an increasing frequency, the single-helical structure
transforms to the half-cylinder, and then reverses to the
single-helical structure. At the larger amplitude, when we
increase the frequency, the phase transition occurs more than
once. In addition to the phase transition from the single-helical
structure to half-cylinder, and then to the single-helical
structure, the phase morphology turns to half-cylinder finally, as
shown in figures~\ref{Fig2}~(e)--(h). Furthermore, we also observe that the
morphology turns directly to the half-cylinder from the
single-helical structure with an increasing amplitude at the
frequency 0.09. At some frequencies, such as
$0.07\leqslant\omega\leqslant0.08$, while increasing the amplitude, the
phase transition occurs from single-helical structure to
half-cylinder, then reverses to a single-helical structure, and
then transforms to a half-cylinder structure. On the whole, the
single-helical structure mainly concentrates on the region of the
weak oscillatory shear flow, which can be seen in the bottom
left-hand corner in figure~\ref{Fig1}; the half-cylinder structure mainly
concentrates on the region of the strong oscillatory shear flow,
which can be seen in the top right-hand corner in figure~\ref{Fig1}.
Moreover, the critical shear frequency of the phase transition
becomes smaller and smaller with an increasing amplitude.

As seen in figure~\ref{Fig2} mentioned before, it shows the morphology
evolution with an increasing frequency at two typical amplitudes,
$\gamma=0.01$ and $\gamma=0.41$. Phase A is represented by black
regions, phase B by gray regions. In figures~\ref{Fig2}~(a)--(d), no phase
transition occurs during the frequency and in figures~\ref{Fig2}~(e)--(h), the
phase transition occurs more than once. These results are mainly
due to the interplay of two factors: the field effect caused by
the oscillatory shear flow and the confinement effect produced by
the confinement boundary. Meanwhile, the field effect is the
combined effect of both amplitude and frequency. For the
single-helical structure in the bottom left-hand corner, it is the
same as the structure that is confined in the cylinder with no
shear of the same diameter, which is in agreement with the
previous studies by Morita~\cite{b1} and Pinna~\cite{ff}. They found
that the helical structure was usually observed at incommensurate
conditions. In our simulation, the length and diameter are
slightly incommensurable with the domain spacing; the amplitude
and frequency are relatively low, the field effect is weaker while
the confinement effect is dominant. Thus, similarly, the
single-helical structure is due to the extensional forces when the
length, diameter and the lamellae spacing are not commensurate.
For the half-cylinder structure in the top right-hand corner, the
amplitude and frequency are higher. Relatively speaking, the field
effect plays an important role, so we get the half-cylinder
structure along the direction of oscillatory shear flow. For the
repeated phase transition with an increasing frequency at a
certain amplitude in figures~\ref{Fig2}~(e)--(h), we consider that the
coupling of amplitude and frequency plays a certain role. Compared
with the field effect, the confinement effect has an absolute
superiority when the frequency is weaker. It is easier to form a
single-helical structure [figure~\ref{Fig2}~(e)]. With increasing the
frequency, the movement of segments A and B strengthens along the
oscillatory shear direction, the coarse graining process is also
intensified along the field direction and then the ordered
half-cylinder structure is formed [figure~\ref{Fig2}~(f)]. With a further
increase of the frequency, the movement is faster, more and more A
and B segments accumulate, they push each other and then it formes
a single-helical structure again [figure~\ref{Fig2}~(g)], and the nature of
this structure is different from the helical under a weaker shear.
When the frequency is high enough, the field effect plays a
leading role, it forms the ordered half-cylinder structure along
the pipe axis [figure~\ref{Fig2}~(h)].

\begin{figure}[!t]
\vspace{-2mm}
\begin{center}
\includegraphics[width=0.8\textwidth]{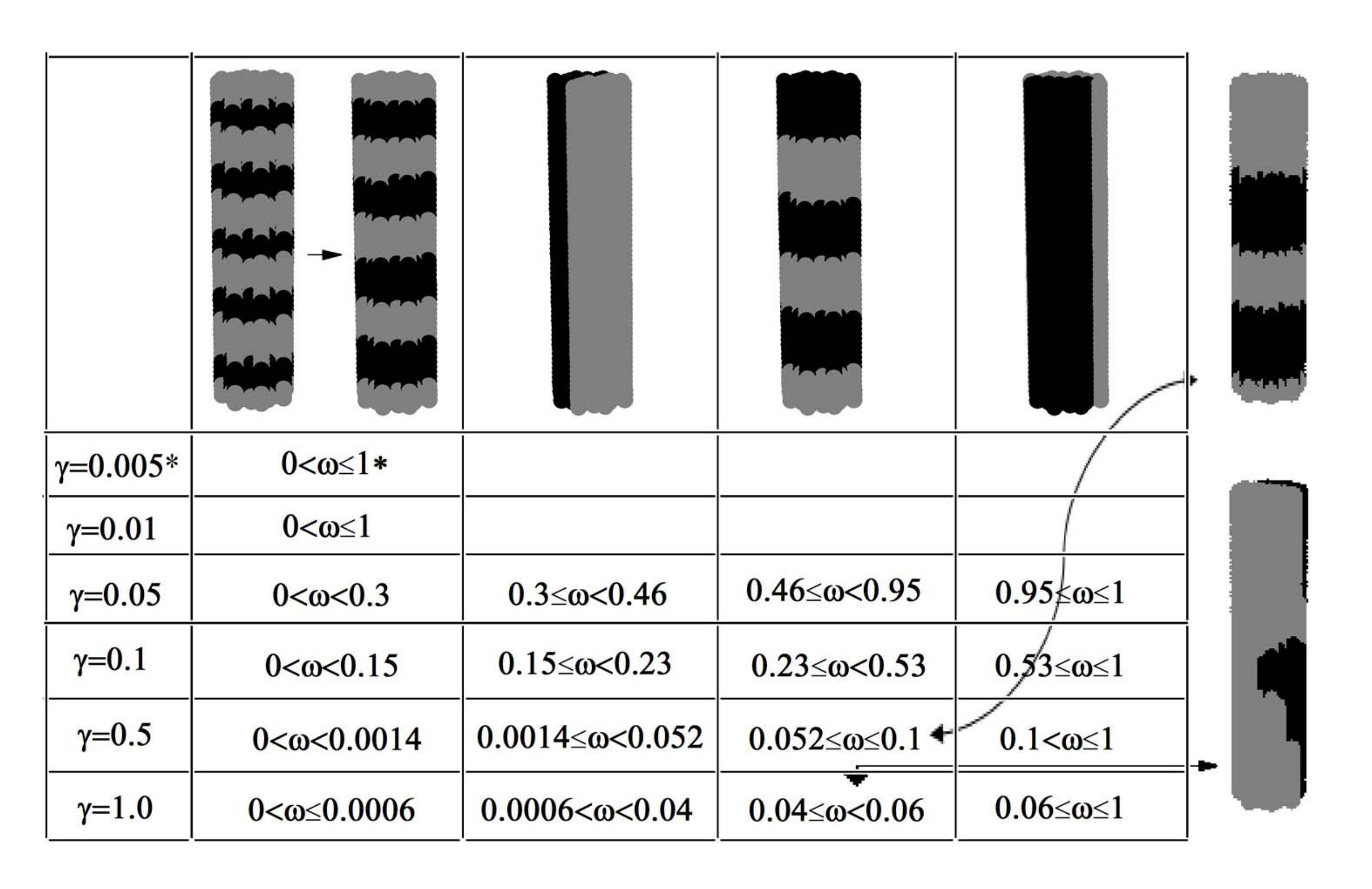}
\end{center}
\vspace{-4mm}
\caption{Simulation snapshots of the diblock copolymer in nanopore $D/L_0=0.76$ under oscillatory shear at different amplitudes $\gamma$ and frequencies $\omega$. Phase A is represented by the black regions, phase B by the gray regions.} \label{Fig3}
\end{figure}

Then, we concentrate on the phase behavior under oscillatory shear
flow in cylindrical nanopore of $D/L_0=0.76$ and
$L_{\mathrm{pore}}/L_0=8.1$. The simulation snapshots under
oscillatory shear flow at different amplitudes $\gamma$ and
frequencies $\omega$ are shown in figure~\ref{Fig3}. As we can see in
figure~\ref{Fig3}, we control the frequency between 0 and 1, and consider
the amplitude of various stages. Overall, it preferentially adopts
a piled-pancakes structure when the shear is low; while it adopts
a half-cylinder structure that is parallel to the pipe axis when
the shear is high. When $\gamma=0.005$ and $\gamma=0.01$, the
domain structure remains the piled-pancakes in the entire
frequency range. However, the number of pancakes decreases with an
increase of frequency when the amplitude $\gamma$ is 0.01, at the
same time, the layer spacing becomes wider. $*$ represents the
morphology and has no change, and it is used to distinguish these
two different cases. When $\gamma=0.05$ and $\gamma=0.1$, with an
increasing frequency, the piled-pancakes structure evolves into
half-cylinder structure due to the aggravation of the coarse
graining process along the pipe axis. At the further increase of
frequency, at the coupling of amplitude and frequency, the system
transforms to the piled-pancakes structure and the number of
pancakes decreases. Undoubtedly, then it turns to the
half-cylinder structure when the shear is strong. It is certain
that the critical shear frequency of phase transition and
frequency range of each structure are different from each other.
This phenomenon also exists in $\gamma=0.5$ and $\gamma=1.0$. The
higher the amplitude is, the lower the critical shear frequency
is. The frequency range in the same structure is smaller when the
amplitude is higher, just like the piled-pancaked in
$0<\omega<0.0014$, $\gamma=0.5$ and $0<\omega\leqslant0.0006$,
$\gamma=1.0$; and the half-cylinder in $0.0014\leqslant\omega<0.052$,
$\gamma=0.5$ and $0.0006<\omega<0.04$, $\gamma=1.0$. Also, the
number of pancakes that formed again are less than before, as
well, the layer spacing is wider than before in $\gamma=0.5$.
However, it forms a mixed structure between piled-pancakes and
half-cylinder just in a small range (i.e., $0.04\leqslant\omega<0.06$)
under $\gamma=1.0$. This is chiefly because at $\gamma=1.0$, the
vibration amplitude along the pipe axis is large, it is difficult
to stack and easy to stretch along the field direction. Certainly,
in the end of the frequency range, the system turns to a stable
half-cylinder structure of paralleling to the pipe axis.

Similar to the reason of the phenomenon in nanopore of
$D/L_0=1.01$, it is the result of the combination of the
confinement effect and field effect. At a smaller amplitude, no
phase transition mainly depends on the confinement effect.
However, the phase transition at a larger amplitude is mainly
decided by the size of the frequency. The slower the frequency is,
the weaker the field effect is. Thus, it is easier to form
piled-pancakes structure, as well as the structure with no shear,
which is in agreement with the study by Pinna~\cite{ff}, as a
result of the competition between the effect of the neutral
surface and the effect of incommensurability. When the shear is
weaker, the surface effect prevails at a small diameter, while
above a certain diameter, the bulk effect takes over. The first
phase transition to the half-cylinder with an increase of the
frequency is due to the intensified graining process along the
pipe axis caused by the strong oscillation of A and B segments. In
case of a further increase of the frequency, these segments would
be more piled up in $z$-axis and jostle each other to form the
piled-pancakes structure due to the too fast vibration. The last
phase transition is caused by a strong field effect, and the
half-cylinder structure is obtained so that A phase and B phase
are completely separated. Furthermore, the decrease of the
critical frequency with an increase of amplitude also indicates
that the phase transition occurs only if the amplitude and
frequency intercouple with each other.

\begin{figure}[!h]
\begin{center}
\includegraphics[width=0.75\textwidth]{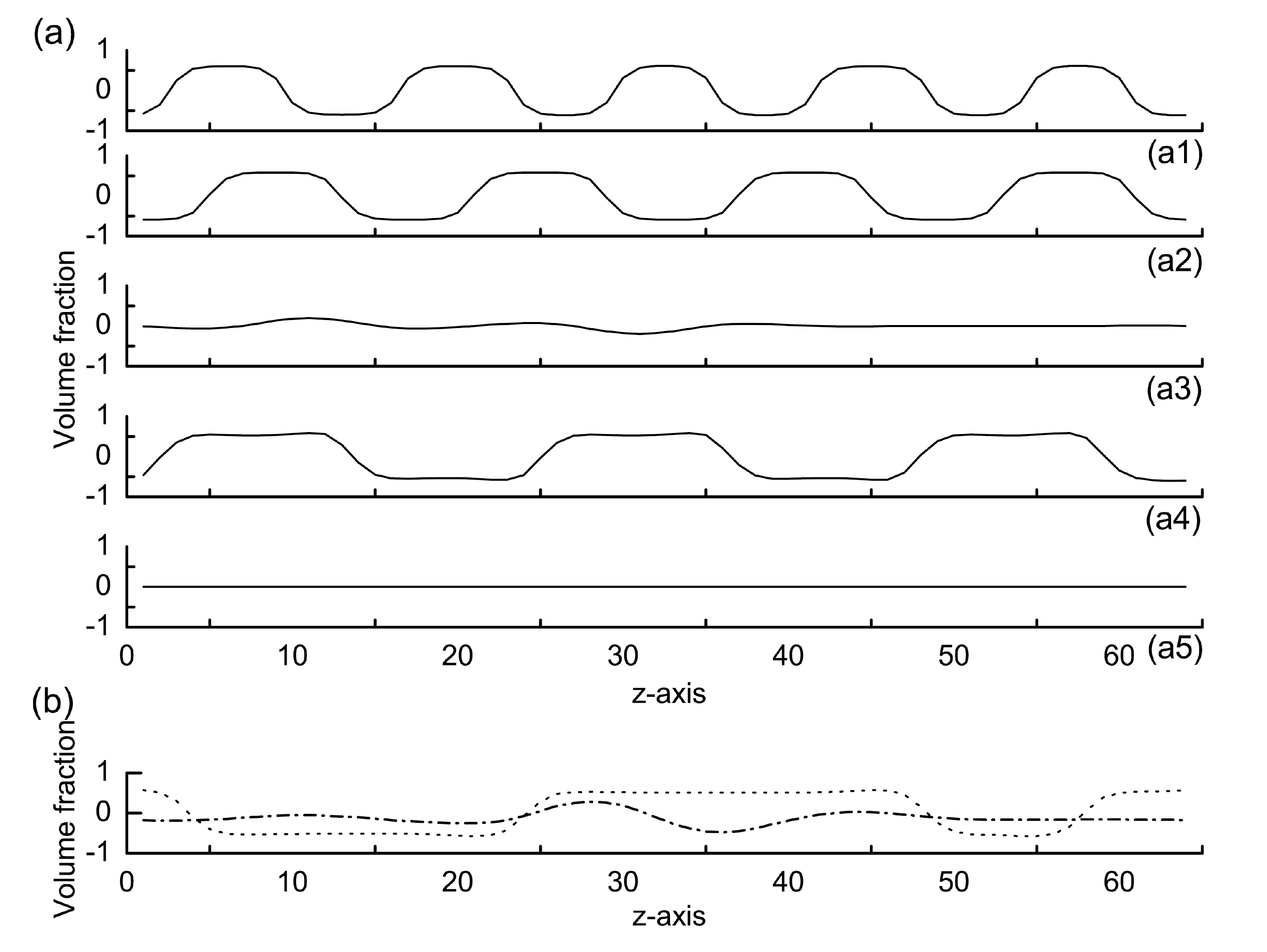}
\end{center}
\vspace{-2mm}
\caption{The order-parameter profiles of $\phi$ along the
pipe axis under different oscillatory shears. (a)~$\gamma=0.1$, from top to bottom, the
frequencies are $\omega=$0.0001, 0.04, 0.15, 0.25, 1.0, respectively. (b)~The dotted line
represents $\gamma=0.5$, $\omega=0.052$; the dash-dotted line represents $\gamma=1.0$, $\omega=0.04$.} \label{Fig4}
\end{figure}

In figure~\ref{Fig4}, the above mentioned phase transition can be clearly
seen in a different way. It displays the order-parameter profiles
of $\phi$ along the pipe axis under different oscillatory shears.
As seen in figure~\ref{Fig4}~(a), $\gamma=0.1$, from (a1) to (a5), the
frequencies are $\omega=$0.0001, 0.04, 0.15, 0.25, 1.0, respectively.
$\phi>0$ represents A-rich domain, $\phi<0$ represents B-rich
domain. This indicates that if the profile of $\phi$ along $z$-axis
has convexities and concavities, the convexity represents A phase,
while the concavity represents B phase. From figures~\ref{Fig4}~(a1) to (a2),
the number of convexities decreases from 5 to 4 with an increase
of frequency from 0.0001 to 0.04 which corresponds to the domain
morphologies in figure~\ref{Fig3}. An approximate straight line in figure~\ref{Fig4}~(a3)
represents $\phi\approx0$, it means that A and B phases are
uniformly distributed in the $z$-axis. Thus, the approximate
straight line corresponds to the half-cylinder structure in
$0.15\leqslant\omega<0.23$, $\gamma=0.1$ as shown in figure~\ref{Fig3}. Then, the
profile of $\phi$ along the $z$-axis changes from a straight line
[figure~\ref{Fig4}~(a3)] to a curved line [figure~\ref{Fig4}~(a4)], and then to a perfect
straight line   [figure~\ref{Fig4}~(a5)] with an increase of frequency from
0.15 to 1.0. It means that the half-cylinder structure caused by
the field effect is more perfect than that caused by the intensity
coarse graining process. In figure~\ref{Fig4}~(b), the dotted line represents
the order-parameter profile of $\gamma=0.5$, $\omega=0.052$, and
the dash-dotted line represents the order-parameter profile of
$\gamma=1.0$, $\omega=0.04$. As we can see, they are completely
consistent with the structures that an arrow points to in figure~\ref{Fig3}.
\begin{figure}[!t]
\begin{center}
\includegraphics[width=0.8\textwidth]{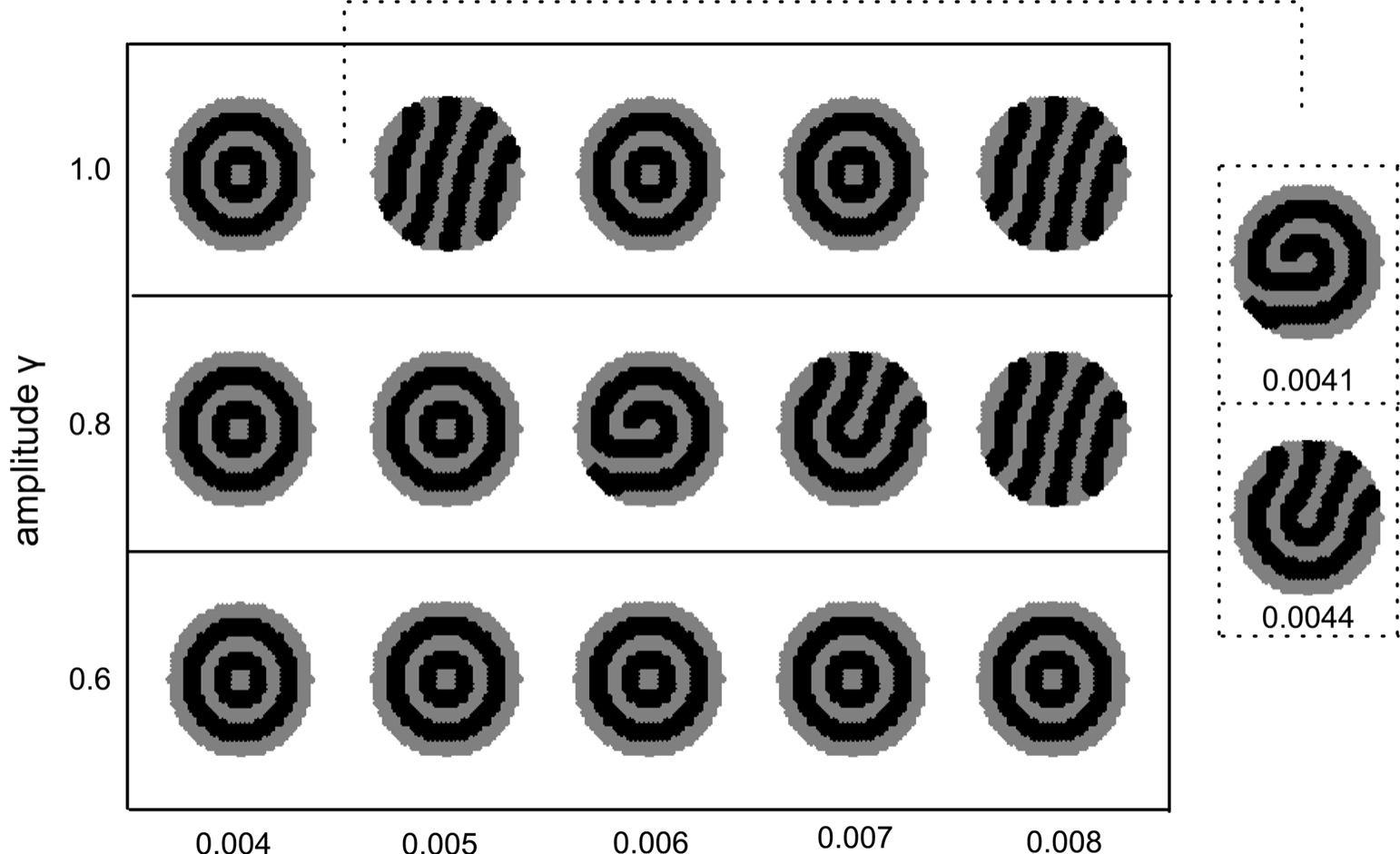}
\end{center}
\caption{The top view phase diagrams of diblock copolymer in
nanopore of $D/L_0=4.05$ under different oscillatory shears. Phase A is represented by the black regions,
phase B by the gray regions.} \label{Fig5}
\end{figure}

\begin{figure}[!b]
\begin{center}
\includegraphics[width=0.6\textwidth]{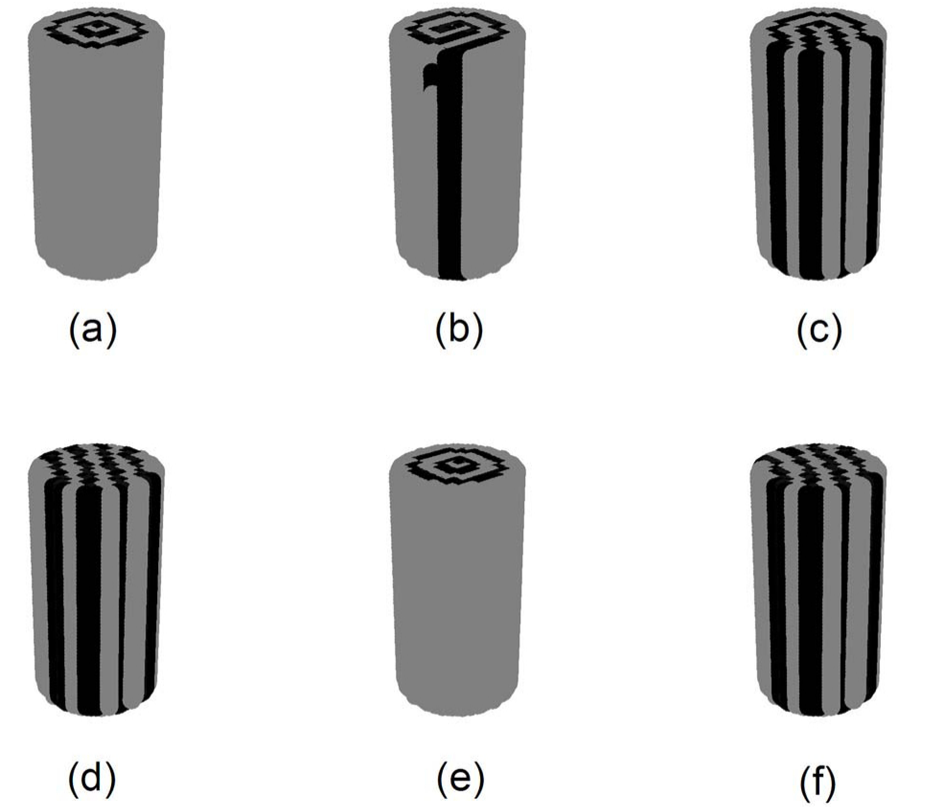}
\end{center}
\caption{Pattern evolution of diblock copolymer in nanopore of $D/L_0=4.05$ under different oscillatory shears: $\gamma=1.0$, (a)~$\omega=0.004$, (b)~$\omega=0.0041$, (c)~$\omega=0.0044$, (d)~$\omega=0.005$, (e)~$\omega=0.006$, (f)~$\omega=0.008$.} \label{Fig6}
\end{figure}

For a relatively large nanopore with $D/L_0=4.05$ and
$L_{\mathrm{pore}}/L_0=8.1$, the phase diagrams of top view under
different oscillatory shears are shown in figure~\ref{Fig5}. The
characteristic morphologies are found mainly in the small
frequency range of $0.004\leqslant\omega\leqslant0.008$ in
$0.6\leqslant\gamma\leqslant1.0$. In general, the most interesting thing is
the big response caused by a small effect, figure~\ref{Fig5} is just this
kind of situation. As we have seen, when $\gamma=1.0$, phase
transition occurs repeatedly in a very small frequency range. The
transformation occurs from the initial concentric ring structure
to the parallel lamellar structure, then the concentric ring comes
into being once again, and then transforms to the parallel
lamellae with an increase of frequency. Thereinto, there exists an
intermediate state in the process of turning to the parallel
lamellae for the first time, it is the ``Swiss roll'' and
``horseshoe'' structures in $\omega=0.0041$ and $\omega=0.0044$,
respectively. At intermediate amplitude, $\gamma=0.8$, since the
amplitude is not enough large, the phase transition only occurs
from the concentric ring to the parallel lamellae. Certainly, the
intermediate state (i.e., the ``Swiss roll'' and ``horseshoe''
structures) also exist. However, when the amplitude decreases to
0.6, we can find that all the structures are a concentric ring in
the given frequency range. Due to the amplitude being relatively
small, even if the frequency increases to 0.008, the field effect
is still weaker than the confinement effect.

As seen in figure~\ref{Fig6}, it shows the morphologies of full view that
corresponds to $\gamma=1.0$ in figure~\ref{Fig5}. The three-dimensional
morphology evolution with an increase of the frequency in figure~\ref{Fig6}
gives us a better visual perception than that in figure~\ref{Fig6}. As for
the reason why this phase transition occurs, we can also interpret
it as the interplay of the field effect caused by the oscillatory
shear and the confinement effect produced by the confinement
boundary. The concentric ring structure is formed under this large
nanopore with no shear, and the same as the structure of the
weaker shears, because the field effect is almost neglected. When
the shear is strong, the field effect prevails, then it forms the
parallel lamellae along the shear axis. The phase transition in
the given frequency range is mainly induced by the amplitude and
frequency intercoupling with each other as before.

\section{Conclusions}\label{Conc}

In the present work, we have combined two different control
measures of the cylindrical confinement and oscillatory shear flow
to manipulate the self-assembly nanostructures and obtained novel
morphologies. Specifically, we have predicted the phase behavior
of diblock copolymer confined in nanopore under oscillatory shear
by considering the different $D/L_0$ and different shears. We have
found that in nanopore of $D/L_0=1.01$, when the amplitude is
small, there is no phase transition observed at a given frequency;
while at a certain amplitude, the system undergoes a phase
transition from single-helical structure to half-cylinder, and
then reverses to a single-helical structure, transforms to
half-cylinder structure with an increase of shear frequency,
finally. When we put the diblock copolymer in nanopore of
$D/L_0=0.76$, the situation is similar, except that the structure
of single-helical structure is replaced by piled-pancakes. When
the diblock copolymer is placed in a large nanopore of
$D/L_0=4.05$, similarly, no phase transition is observed at a
smaller amplitude, and the phase transition occurs with an
increase of the frequency under a relatively bigger amplitude. It
changes from a concentric ring to the parallel lamellar structure,
then reverses to a concentric ring, and finally, turns to a
parallel lamellar structure. However, in the process of phase
transition, there exist two transient states, they are the ``Swiss
roll'' and ``horseshoe'' structures. We have discussed the
transformation of three kinds of typical structures and
constructed a phase diagram of different forms with the changing
amplitude and frequency. Although the morphologies at different
$D/L_0$ are different, the reason for the phase transition with
the change of amplitude and frequency is roughly the same, which
is the interplay of the field effect caused by the oscillatory
shear and the confinement effect produced by the confinement
boundary. These results can provide an easy method to create the
ordered, defect-free nanostructured materials for an
experimentalist through the combined control measures of the
cylindrical confinement and oscillatory shear flow.

\section*{Acknowledgements}

Project supported by the National Natural Science Foundation of
China (Grant No.~21373131), \linebreak the Specialized Research Fund for the
Doctoral Program of Higher Education of China (Grant \linebreak
No.~20121404110004), and the Provincial Natural Science Foundation
of Shanxi (Grant No.~2015011004), the Research Foundation for
Excellent Talents of Shanxi Provincial Department of Human
Resources and Social Security.

\newpage

\ukrainianpart
\title
{Циліндрично обмежене скупчення  діблочного кополімера під дією  осциляційного зсувного потоку}

\author
{Ю.-К.~Гуо, Ю.-К.~Пен, Дж.-Дж.~Жанг, М.-Н.~Сун, В.-Ф.~Вонг, Х.-Ш.~Ву}

\address{Школа хімії та матеріалознавства, Шанхайський нормальний університет, Лінфен, 041004, Китай}

\makeukrtitle

\begin{abstract}
Керування самоскупченими наноструктурами  з допомогою поєднання різних заходів керування виявляється багатообіцяючим шляхом для чисельних застосувань для генерування шаблонів  для наноструктурованих матеріалів.
У цій статті двома різними заходами керування  є циліндричне обмеження та осциляційний зсувний потік. Досліджується фазова поведінка діблочного
кополімера, обмеженого нанопорою під дією осциляційного зсуву з врахуванням різних $D/L_0$ ($D$~--- це діаметр циліндричної пори, $L_0$~--- це розмір домена) та різних зсувів  з допомогою симуляцій коміркової динаміки. При різних $D/L_0\,$, в системі відбувається еволюція різної морфології і фазовий перехід зі зміною амплітуди і частоти.
Тим часом, утворюється ряд нових морфологій. Для кожного  $D/L_0\,$, ми будуємо фазову діаграму різних форм та аналізуємо причини здійснення фазового переходу.  Встановлено, що, хоча морфології відрізняються при різних  $D/L_0\,$, причина фазового переходу при зміні амплітуди і частоти є приблизно однаковою, а саме вона полягає в поєднанні польового ефекту і ефекту просторового обмеження.
Ці результати можуть дати експериментаторам простий метод створення впорядкованих, бездефектних наноструктурованих матеріалів за рахунок поєднання циліндричного обмеження  та осциляційного зсувного потоку.
\keywords самоскупчення, блочний кополімер, циліндричне обмеження, осциляційний зсувний потік

\end{abstract}

\end{document}